\documentclass[%
 amsmath,amssymb,
 aps,
prfluids
]{revtex4-2}

\usepackage{graphicx}
\usepackage{dcolumn}
\usepackage{bm}

\newcommand{\appname}{Appendix } 
\newcommand{\figname}[1]{Fig.~{#1}} 
\newcommand{\eqname}[1]{Eq. ({#1})} 
\newcommand{\ie}{\textit{i.e.}} 
\newcommand{\eg}{\textit{e.g.}} 
\newcommand{\manuscriptypename}{letter} 
\newcommand{\figuresectionname}{} 
\newcommand{\datafoldername}{.} 
\newcommand{\partorchapterrname}[1]{} 
\newcommand{\citname}[1]{~\cite{#1}} 
\newcommand{\citpname}[1]{~\citep{#1}} 

\begin{document}

\preprint{APS/123-QED}

\title{The characteristic rupture height of the mediating air film beneath an impacting drop on atomically smooth mica}

\author{Ramin Kaviani}
\author{John M. Kolinski}%
 \email{john.kolinski@epfl.ch}
\affiliation{%
EMSI Laboratory, {\'E}cole Polytechnique F{\'e}d{\'e}rale de Lausanne, Switzerland
}%




\date{\today}

\begin{abstract}
Before a droplet can contact a surface during impact, it must first displace the air beneath it. Over a wide range of impact velocities, the droplet first compresses the air into a thin film, enhancing its resistance to drainage; this slows the progress of the liquid toward the surface. Indeed, below a critical impact velocity, the air film remains intact, and rebounds off of the air film without making contact. For impact velocities exceeding this critical impact velocity, the droplet always makes contact. The initiation of contact formation requires a topological transition, whereby the initially connected gas domain is ruptured, and a liquid capillary bridge forms, binding the droplet to the surface. Here we probe this transition in detail around the critical impact velocity using calibrated total internal reflection (TIR) microscopy to monitor the air film thickness and profile at high speed during the impact process. Two air film rupture modalities are observed: nucleated contacts, which are isolated and don't correspond to the global minimum air film thickness, and spontaneous contacts, that occur always on a ring centered upon the impact axis where the air film reaches its global minimum. Our measurements show that for impact velocities exceeding the critical velocity for contact initiation, the air film ruptures at a nearly identical height $h_{\min} \approx 20$ nm, for two fluids - silicone oil and a water glycerol mixture. The height and time duration of the air film prior to contact, analyzed for over 180 droplet impact experiments, are consistent with a linear instability driven by van der Waals forces in the range of experimentally measured values of the Hamaker constant. Impact events of water solution droplets show different statistics for contact nucleation than the silicon oil; this suggests another mechanism may dominate contact nucleation during impact of the solution; nevertheless, a critical impact velocity above which contact always occurs is identifiable for both liquids. 
\end{abstract}

\maketitle


\section{Introduction}
\label{s_Intro}
The initial formation of contact between a liquid and a solid surface is typically mediated by a third medium - most often in our daily experience, air. Indeed, when a droplet of water impacts a solid, it makes contact imperceptibly quickly - a transition so rapid that it occurs much faster than we can observe with the unaided eye. However, from a continuum perspective, we know the gas must be punctured, or otherwise disrupted, for liquid-solid contact to form. The process of droplet impact is at the center not only of our daily experience, but also in myriad industrial settings such as heat transfer, fuel injection and inkjet printing\citpname{Review:Josserand2016}. The role of air in droplet impacts has been thoroughly studied, from its initial formation and evolution\cite{Droplet:Mandre2009, Droplet:Kolinski2012, Droplet:VanderVeen2012, Droplet:Bouwhuis2012, Droplet:deRuiter2012, Droplet:Mandre2012, Droplet:Kolinski2014, Droplet:Mishra2022}, non-contact bouncing transition\cite{Droplet:Kolinski2014+, DynamicContact:deRuiter2014, Droplet:Chubynsky2020} to post-contact wetting\cite{Droplet:Thoroddsen2010, lo2017mechanism, Droplet:Kolinski2019}. It is clear that over a wide range of impact velocities, despite the ubiquity and applicability of droplet impact mechanics, the key phase of air film disruption and the formation of contact remains to be precisely quantified. 

Although the air film formation and rupture phenomenon have been identified in prior work, and it is clear that the air film thickness evolves relatively slowly\cite{Droplet:Kolinski2014}, there are striking discrepancies in the value of the air film thickness immediately prior to contact that are reported by different experimental studies\citpname{Droplet:Kolinski2012,DynamicContact:deRuiter2015,Droplet:Lo2017,Droplet:Gao2019} ranging from $1-2$ nm to several hundred nm. With this large spread in the measured air film thickness prior to rupture, the mechanism responsible for the instability of the air film has not yet been conclusively determined. Indeed, various mechanisms ranging from spinodal dewetting\cite{Droplet:Wyart1990, reiter1999thin, Droplet:Kolinski2012} to gas kinetic effects\citpname{Droplet:Chubynsky2020} to electrostatics\cite{Droplet:Gao2019+} or surface roughness\cite{latka2012creation, Droplet:Kolinski2015} have been identified as possible causes for air film rupture; however, the spread in the data of the measured air film thickness prior to rupture makes it impossible to distinguish between the various mechanisms that may dominate these dynamics. Thus, a fundamental and important question of what drives liquids into contact with solid surfaces through an intervening gas remains open, with implications for coating flows\cite{weinstein2004coating} and splashing\cite{Droplet:Xu2005}.

In this letter, we present measurements of the temporal evolution of the air film that are calibrated to $+7/-3$ nanometers. To achieve this calibration, we employ two optical methods that are used simultaneously to ensure rigorous calibration of the measurement - total internal reflection microscopy (TIR)\citpname{Droplet:Kolinski2012,TIR:Shirota2017} and single-color interferometry (FIF)\cite{driscoll2011ultrafast}. Once the simultaneous imaging yields a calibrated TIR measurement with a prescribed displacement of a precision piezo stage, we use the higher precision TIR measurement to image the liquid-air interface during droplet impact over a velocity range from $0.3$ to $0.9$ m/s on freshly-cleaved, atomically smooth mica substrates at high-speeds (80,000 fps, Photron Nova S12). This velocity range contains the rebound-to-wetting transition\citpname{Droplet:Kolinski2014+} for both water-glycerol and silicon oil droplets. To control the impact velocity precisely, droplets are released from a luer tip, with a diameter selected to yield identically sized droplets of either water-glycerol or silicon oil. To establish good statistics of the impact dynamics, multiple impacts were carried out at each impact velocity. Analysis of the recorded film profiles yields the height of closest approach prior to rebound, $h^*$, obtained by azimuthally averaging the data for a single event, or the height of closest approach, obtained at each pixel where air film rupture is observed $h_{\min}$. 

Despite the great care we take in preparing the mica surfaces, nucleation of contact is nevertheless occassionally observed, oftener for the water-glycerol solution than for the silicon oil; the time to wait between air film formation and eventual rupture, $t_c$, is recorded for each event. The $t_c (h_{\min})$ data are then analyzed using the prediction of linear stability analysis for the timescale for the development of the linear instability. A wave-number and Hamaker constant pair yield a line that cleanly separates the nucleated contacts from the spontaneous rupture events, highlighting the consistency of the observed air film destabilization events with the interfacial-forces driven instability via the spinodal dewetting of the air film. We comment on how this observation might prove useful in future numerical calculations, and its importance to liquid-solid contact formation more generally.  


\begin{figure}[!hpt]
 \centering
\includegraphics[width=0.9\textwidth]{\datafoldername/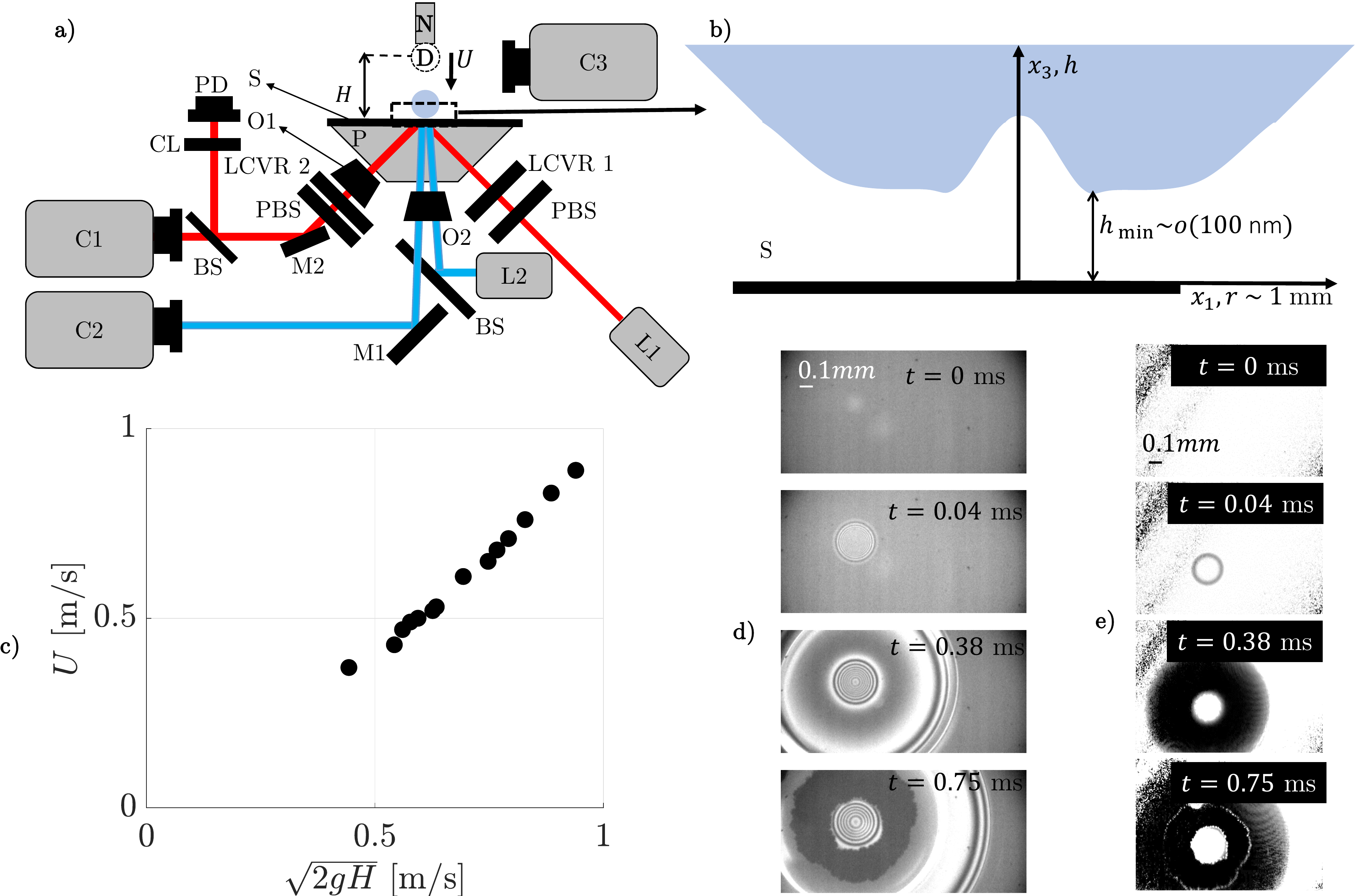}
\caption{
a) Schematic layout of simultaneous TIR and FIF microscopy for the study of droplet impact. The setup is calibrated to have an accuracy of $\pm 5$ nm for air layer thickness measurements below $300$ nm in the $x_3$ direction (see \appname\ref{s_Calibration}).   
The liquid drops with diameter ($D \approx 2$ mm) fall from the height $H$ from the nozzle (N) on a mica substrate (S) reaching maximum velocity of $U$. We cleave a mica slab through its thickness to obtain an atomically flat surface over several  millimetres in $x_1-x_2$ plane. We optically couple the mica sheets to the dove prism (P) made from BK-7 glass by immersion oil. 
In the TIR microscopy optical path, coherent light emitted by the laser source (L1) illuminates S at incident angles greater than the critical angle for the solid-air interface. We use the polarizing beam splitters (PBS) and liquid-crystal variable retarders (LCVR) to ensure that the light arrives at the substrate-air interface linearly p-polarized. 
The light is totally internally reflected from the mica-air interface beneath the impacting droplet. Half of the reflected light is captured by camera C1 and the second half is concentrated on a photo diode senor (PD) by a condenser lens (CL). PD is required for measuring the optical properties of substrate S as explained in \appname\ref{s_Calibration}.  
FIF data are acquired using a second optical path. Light emitted from a high power LED light source (L2) passes through the beam splitter (BS), reflects from the impact interface, and is captured on camera (C2). 
Two dry 5X Mitutoyo microscope objectives (O1 and O2) are used for magnification.
The monochromatic fast cameras C1 and C2 are fitted with tube lenses (not shown), and are synchronized to record images at $80,000$ FPS.
Mirrors (M1 and M2) are used to align the light.
We use a third camera (C3) recording at $8,000$ FPS to measure $U$.
b) The region underneath the droplet just before contact initiation is shown schematically with an exaggerated $x_3$ axis: the air forms a nanometer-scale lubrication layer, at the periphery of a dimple centered upon the impact axis.
c) Impact velocities, $U$, ranging from $0.34-0.91$ m/s, are controlled by changing $H$. d)-e) Example image series of the air film formation and rupture as captured with the FIF and TIR imaging modalities are shown at three time steps. After post-processing, the light intensities can be converted to the air film thicknesses for each image; the temporal evolution is recovered by evaluating the time series of images. The dimple region centered upon the impact axis of the droplet is clearly visible; it extends a few microns from the impact surface, as can be seen from the fringe rings in the FIF images. This region resides outside the penetration depth of the TIR imaging modality, and is identifiable from the bright circles centered upon the impact axis in the TIR images.} 
\label{f_DropletSetup}
\end{figure}

\section{Experimental Setup and Methods}
\label{s_DropletSetup}

The experimental setup consists of two distinct optical paths corresponding to the TIR and FIF experimental methods used to characterize the liquid-air interface beneath the impacting drops. The first path, dedicated to the TIR measurement, directs coherent laser light to strike the impact surface obliquely from below, whereas the second path, dedicated to the FIF measurement, directs light from a weakly coherent ($\ell_{coh} \approx$ 2.5 $
\mu$m such that it is normally incident upon the impact surface from below; the key components for each optical path are depicted schematically in 
\figname{\ref{f_DropletSetup}(a)}. Such simultaneous profilometry provides complementary data for interfacial mechanics\citpname{Droplet:Pack2017,Droplet:Pack2018,Neural:Chantelot2021,Neural:Kaviani2022}. The liquid-air interface prior to contact initiation is depicted schematically in \figname{\ref{f_DropletSetup}(b)}. To control the impact velocity $U$, we vary the height $H$ from which the droplet is allowed to fall; due to the residual oscillations of the droplet prior to impact, a weak non-linearity of $U$ emerges as a function of $H$, as shown in \figname{\ref{f_DropletSetup}(c)}. Two example time-synchronized image series, recorded with the two respective optical paths, are depicted in \figname{\ref{f_DropletSetup}(d) \& (e)}. 

To probe the stability of the air film, the defects commonly encountered on even carefully cleaned glass surfaces must be avoided. This requires a solid surface that will be devoid of material defects over an area of at least ca. 1 mm$^2$. To obtain such `perfect' surfaces, we cleave sheets of high-grade mica through their thickness, generating an atomically smooth surface over several mm$^2$. Experiments are conducted with two different liquids: water-glycerol solution droplets with $D=1.91$ mm and silicon oil with $D=1.83$ mm. We used silicon oil with a viscosity of $10$ cSt and prepared the water-glycerol solution mixture to have the same kinematic viscosity. The density of the solution and oil are $1168.1$ kg/m\textsuperscript{3} and $937.1$ kg/m\textsuperscript{3}, respectively. The surface tension for the solution ($68.0$ mPa.m) is greater than of the oil ($20.2$ mPa.m) as measured by pendant drop method. Both liquids wet the mica surface with immeasurably small contact angles. The finite viscosity of the liquid is used to suppress high-frequency oscillations that can rupture the air film upon the center of the impact axis\cite{zhang2021thin}, and suppress droplet rebound. 

Mica is a weakly birefringent solid, and can alter the polarization of the light used in TIR optical path. The birefringence of the mica can be pre- and post-compensated with liquid crystal variable retarder plates, ensuring strict control over the state of polarization at the reflecting surface, where the droplet will impact the surface. While prior studies suggest an analytical means of calibrating the transfer function from intensity to height\citpname{TIR:Shirota2017, Droplet:Kolinski2019}, we used complementary optical methods to perform a direct calibration using a precision piezo stage, as described in detail in \appname\ref{s_Calibration}. The concurrent imaging with TIR and FIF provides complimentary information about the liquid-air-solid interface - TIR can resolve air layer thicknesses up to $\approx 500$ nm, whereas FIF can visualize the air film profile of up to several microns\citpname{FIF:Radler1992}. In FIF, the film profile is encoded in a two-dimensional fringe pattern, where the intensity of the fringes indicates the optical interference at a given spatial location within the image. While the FIF fringe patterns can be demodulated to provide absolute film measurements that are up to few microns thick, we only analyze the regions where the film thickness is less than half of the light wavelength, or approximately $200$ nm.

Our microscopy setup has a resolution of $3.6~\mu$m in the $x_1-x_2$ plane. In the $x_3$ direction, the resolution of profilometry is as low as $7$ nm but the calibrated field of view has a ceiling at about $350$ nm. The threshold of unstable air gap thicknesses ($h^{*}$ and $h_{\min}$) fall within the calibrated range of our imaging modalities. As the dynamics of droplet rebound and air film rupture are very rapid, high-speed imaging is employed to attain relevant temporal information from our experiments. Light from the two optical paths is imaged onto the sensors of synchronized high-speed cameras (Photron Nova S12), and the sensor data is recorded at $80,000$ frames per second. The sequence of images in \figname{\ref{f_DropletSetup}(d) \& (e)}  shows the rapid nature of air film formation, its rupture, and surface wetting.

\begin{figure}[!ht]
 \centering
\includegraphics[width=0.9\textwidth]{\datafoldername/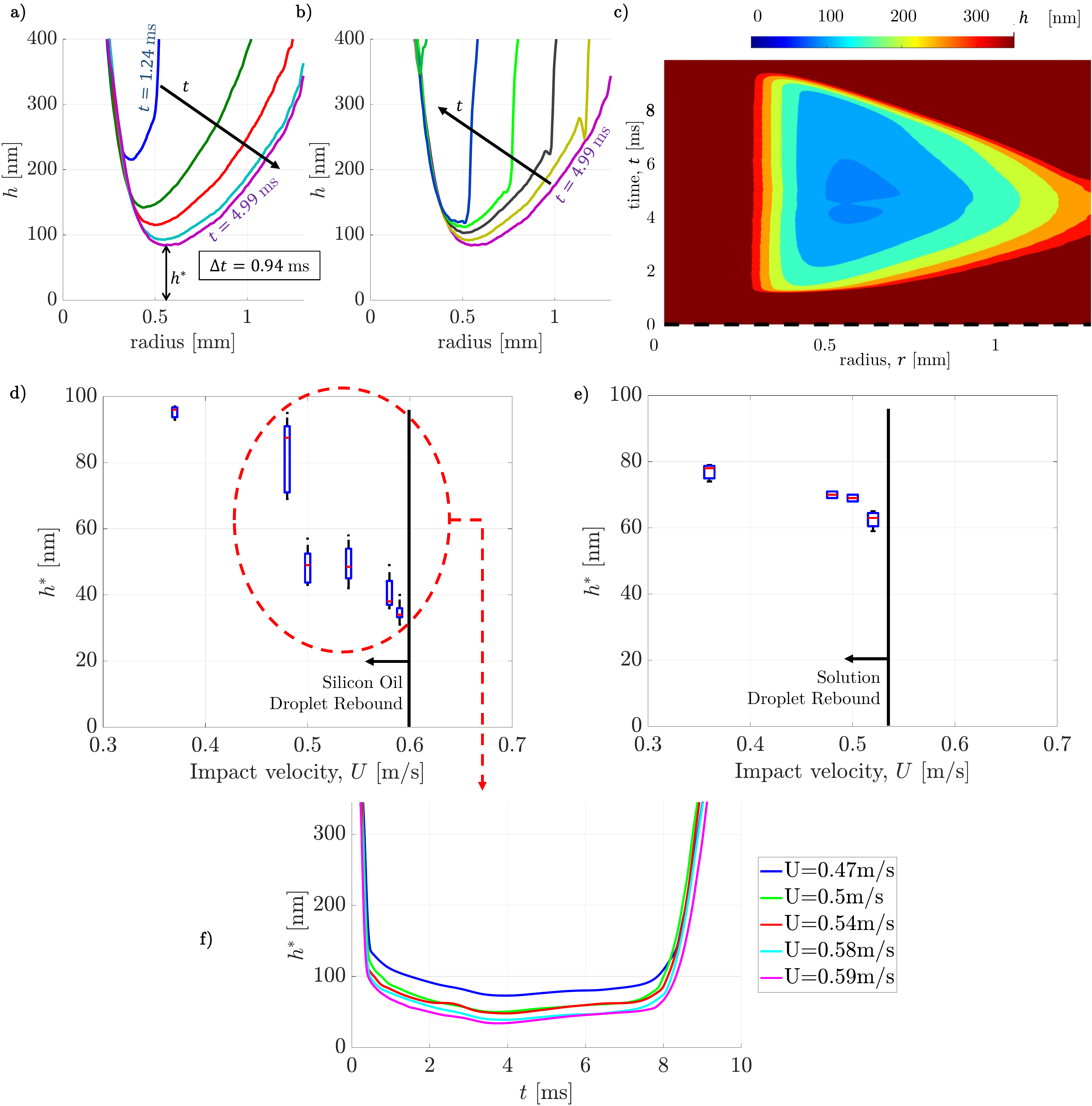}
\caption{Less than a critical impact speed, the air film under the droplet is stable and functions as a cushion that causes the droplet to rebound off the mica substrate. As an example for $U=0.5$ m/s, the $10$ cS water-glycerol droplet pushes the air layer creating a time dependent profile $h$ which is drawn based on the radius, $r$, from the center of the impact in (a-b) for few time steps. The data is truncated at the maximum gap thickness of $350$ nm which is the limit that our microscopy setup is calibrated for. Time ($t$) is measured from the moment time droplet enters the FIF field of view $h(0,t=0)=O(10^{-6})$ m.  Same data is plotted in color map of (c) indicated by the color-bar for continuous $t$. For impact velocities more than a threshold, the air layer is always unstable and the wetting initiates. This indicates a limiting stable film height ($h^{*}$) which decreases gradually with higher velocities as shown in (d,e) for the silicon oil and the water-glycerol solutions, respectively. The bouncing critical velocity  for the oil droplets ($0.6$ m/s) is higher compared to that of the solution ($0.53$ m/s). The $h^{*}(t)$ variation close to critical impact velocity is presented in (f) for the solution. The minimum captured $h^{*}$ for the oil, at $32$ nm, is lower than the solution, at $59$ nm. 
} 
\label{f_Bouncing}
\end{figure}

\begin{figure}[!ht]
 \centering
\includegraphics[width=0.95\textwidth]{\datafoldername/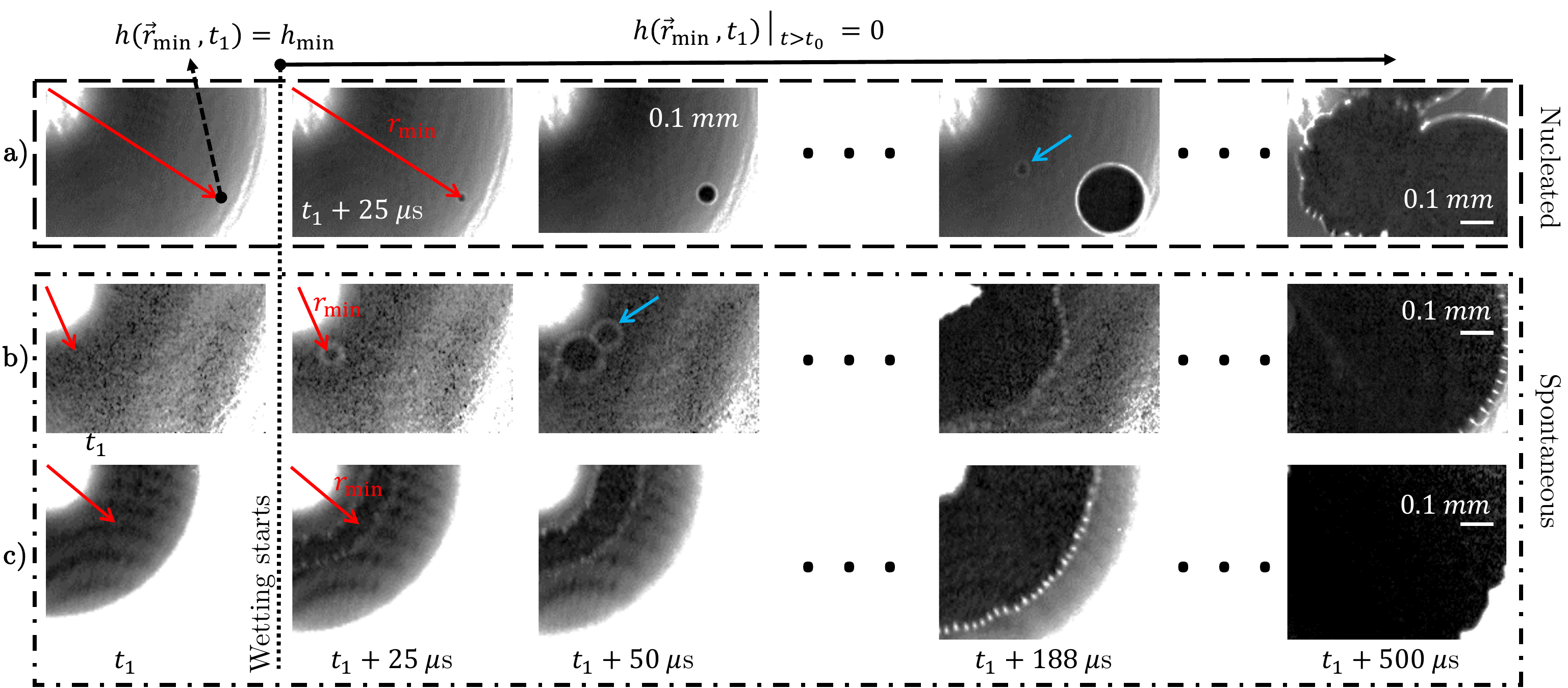}
\caption{
The normalized image series of contact formation from the TIR microscpy for a) $U=0.5$ m/s, b) $U=0.58$ m/s, c) $U=0.91$ m/s. The wetting starts by breaking through a thin layer of air between a mica surface and the liquid. $h_{\min}$ is recorded locally at $r_{\min}$ at the last frame recorded ($t=t_1$) just before the air film rupture is captured by the high speed camera. In (a), 2 nucleated contact form as separate entities and the contact lines moves radially outwards from the respective points of formation until they meet. The spontaneous contact initiation shown (b-c) are considerably more axisymmetric with respect to the center of the droplet ($r=0$).  }
\label{f_ContactFormation}
\end{figure}

\begin{figure}[!ht]
 \centering
\includegraphics[width=0.9\textwidth]{\datafoldername/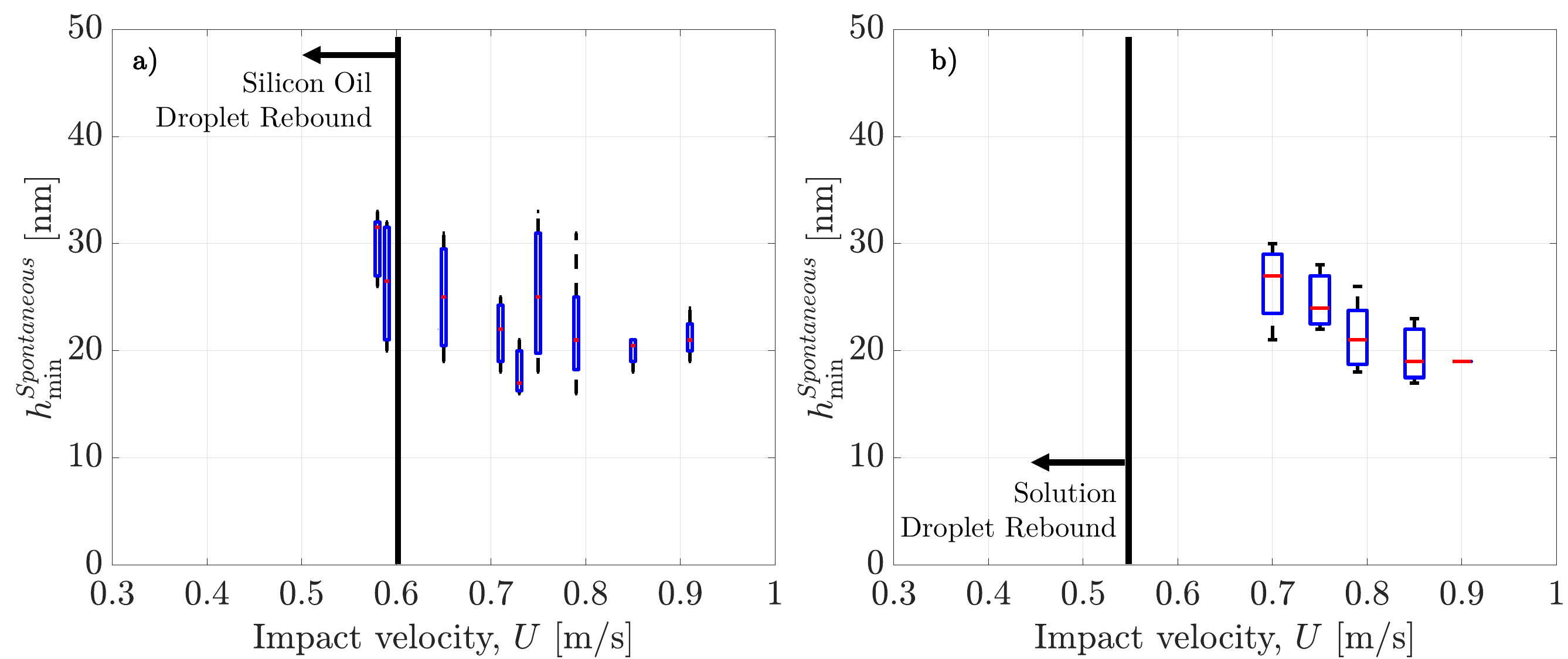}
\caption{
The lowest film height, $h_{\min}$, is plotted for spontaneous contacts in (a,b) for 36 oil impact events and 18 solution impact events, respectively. Spontaneous contact formations are much more frequent for the oil but $h_{\min}$ lies in a range of $18-35$ nm for both liquids. The resolution of the data is limited by the cameras speed and the lower bound of $h_{\min}$ data is the relevant quantity, \ie, $h_{\min}=18\pm3$ nm for both the liquids. Minimum $h_{\min}$'s do not change with the impact speed.  This is evident for the silicon oil. The data density for the solution is less dense for spontaneous impacts, still any changes observed ($\sim 7$ nm) in $h_{\min}^{Spontaneous}$ are within the tolerance of the profilometry measurement. Despite several tries, we did not observe any spontaneous contact formation for moderate velocities of water-glycerol droplet impact, $U < 0.7$ m/s. } 
\label{f_MinHeight}
\end{figure}

\section{Air Film Formation and Rupture} 
At the early stage of a low speed impact, the liquid skates over the air\citpname{Droplet:Kolinski2012}, increasing the lateral extent of the air film. In this regime, the droplet entrains the air at the leading edge of the drop, and depending on the liquid viscosity, can lift off away from the surface\citpname{Droplet:Kolinski2014}, generating the extended liquid-gas interface above the solid. The extended air film drains very slowly, and remains stable provided it doesn't approach within a critical distance of the surface\citpname{Droplet:Kolinski2014+}. Indeed, for sufficiently slow impacts, the droplet will reliably rebound from the air film\citpname{Droplet:Kolinski2014+}, as shown in \figname{\ref{f_Bouncing}(a) \& (b)} . This indicates a threshold height of closest approach, $h^{*}$, that decreases as the impact speed increases (\figname{\ref{f_Bouncing}(c) \& (d)}). The bouncing to wetting transition velocity for the oil droplets ($0.6$ m/s) is higher than that of the solution ($0.53$ m/s). At the same time, the silicon oil droplet reliably approaches closer to the surface: $h^{*} = 32\pm3$ nm for the oil, while $h^{*} = 60\pm3$ nm for the solution.   

The initial formation of contact - via a topological transition from a coherent air film to a locally ruptured air film - is observed to occur in two distinct manifestations. First, and most commonly, the air film ruptures locally, at an isolated contact, \emph{independently} of the relative location of the globally minimal air film thickness. Such events clearly nucleate at a surface imperfection, owing to the proximity of the liquid to the solid over the extent of the thin air film. The outward-advancing wetting front remains symmetrically centered upon the location of initial contact formation as previously described\citpname{Droplet:Kolinski2019}; an example of such a nucleated contact is given in the time series shown in \figname{\ref{f_ContactFormation}(a)}. The second mode of contact formation is typified by the near exact correspondence between the location of air film rupture and the globally minimum air film height. Due to the axisymmetry of the air film profile, such events are often clustered closely in time, and occur at a nearly identical radial distance from the impact axis. Two examples of this spontaneous mode of contact formation are shown in \figname{\ref{f_ContactFormation}(b) \& (c)}. 

For all contacts, we measure the local height of the air film immediately prior to contact formation, $h_{\min}$, using the last frame recorded. $h_{\min}$ is plotted as a function of impact velocity in \figname{\ref{f_MinHeight}}. The $h_{\min}$ distribution for nucleated contacts is uniform with respect to the impact velocity, as described in detail in \appname\ref{s_Nucleation}, where the distribution of contact heights is skewed slightly higher for the water-glycerol solution than for the silicon oil droplet impacts.

In the absence of nucleation, spontaneous contact formation was observed only for $h_{\min} < 35$ nm. Below this height, contact initiates suddenly, forming a capillary bridge that binds the liquid to the solid surface. Notably, for higher velocity impacts, the resolution of $h_{\min}$ can be limited by the acquisition rate, as the drainage rate of the air film increases with larger $U$; thus, depending on the phase of the image acquisition and the contact formation, a different value of $h_{\min}$ might be recorded. For each value of $U$, we repeated experiments several times, and thus serendipitously capture lower values of $h_{\min}$, where the image recorded corresponds to the time immediately prior to contact. In this way, the lower bound of $h_{\min}$ data is the relevant quantity. $h_{\min} = 18\pm3$ nm for both the silicone oil and the water-glycerol solution. Notably, spontaneous contact occurred much more frequently for the oil than for the water glycerol solution - this occurs because nucleation is sporadically observed to occur at larger $h_{\min}$ for the water glycerol, as described in detail in \appname\ref{s_Nucleation}. For all spontaneous contacts, $h_{\min}$ falls in a range of $18-30$ nm for both liquids. $h_{\min}$ for spontaneous contact doesn't change considerably with the impact speed. The changes observed are within the tolerance of our measurement accuracy, at ca. 7 nm.

\begin{figure}[!ht]
 \centering
\includegraphics[width=0.75\textwidth]{\datafoldername/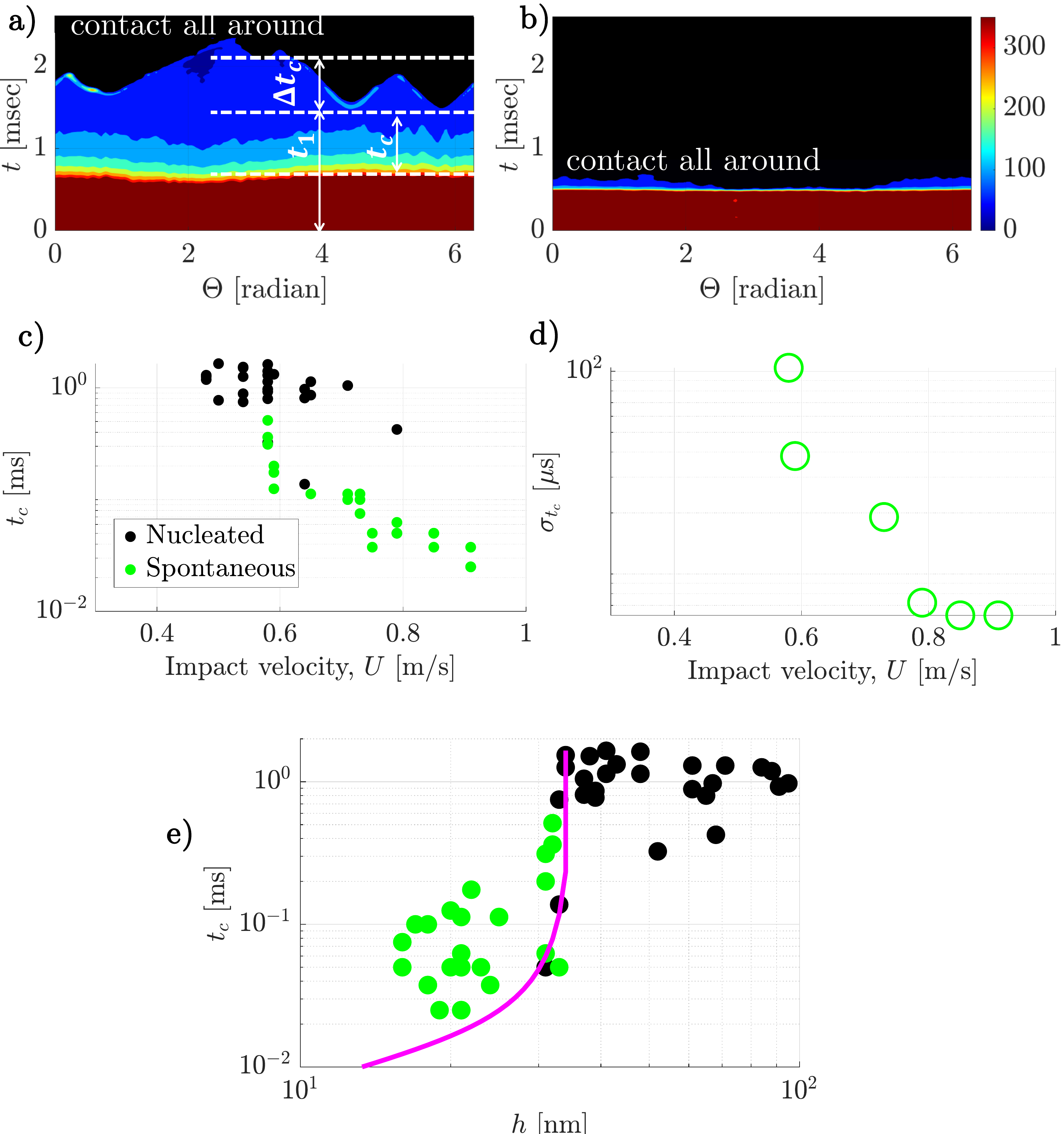}
\caption{
Kymographs of the initiation of contact on the circumference of the ring of contact $r=r_{\min}$, centered on the impact axis are plotted in (a,b) for silicon oil droplet impact velocities of $U=0.58$ and $0.91$ m/s, respectively. We made the $\Theta-t$ kymographs by unwrapping the ring of contact formation, $h(r_{\min},t)$. 
We measured the time $t_c$ between first passage of the liquid above the ring of contact until contact formations. At highest velocity of $0.91$ m/s, the spontaneous contact ring formation along $\Theta=0-2\pi$ starts almost simultaneously. The dynamics occurs close to the limit of camera recording frequency which corresponds to time steps of $12.5~\mu$s as shown in (b). 
Timescale for the development of instability of the air layer under the silicon oil droplets are marked in (a). The data comprise  the pairs of $(U, t_c)$ in (c) and $(h_{\min}, t_c$) in (e). $h_{\min}$'s are measured immediately before the initiation of contact for 53 impact events over the range of $U$. The standard deviation of  $t_c$, defined by $\sigma_{t_c} =\frac{1}{N-1} \sqrt{\sum_{n=1}^{N-1} \Delta t_{c_n}}$ for $N$ number of spontaneous contacts decreases suddenly as $U$ increases as shown in (d). 
The accuracy of $t_c$ is constrained by the camera FPS.  
The magenta line in (e) specifies the transition to wetting based on linear stability analysis (\appname\ref{s_Stability} \& \eqname{\ref{e_stability}}) with estimated $A=3e-19$ J and $\frac{k}{2 \pi}=2e5$ 1/m.
}
\label{f_ContactTime}
\end{figure}

Analysis of the nucleated and spontaneously-formed contacts can provide insight into the mechanisms responsible for air film instability and the transition to contact. Following analysis of prior measurements\citpname{Droplet:Kolinski2015} made without careful calibration as described in \appname\ref{s_Calibration}, we evaluated the statistics of $h_{\min}$ as well as the time from when the air film first forms to contact initiation, $t_c$. The dynamics of air film rupture are depicted graphically for two impact velocities, $U$ = 0.58 and 0.91 m/sec, in \figname{\ref{f_ContactTime}(a) \& (b)}. For $U$ slightly greater than the critical velocity for droplet rebound, the air film is perforated discretely, with capillary bridges formed from the liquid closest to the surface; however, there can be a significant delay between the first air film rupture event and the final air film rupture event, $\Delta t_C$. For much higher velocity impacts, contacts appear to form much more quickly, so that the air film appears to rupture symmetrically from the liquid closest to the surface, within the time resolution of our imaging. As $U$ increases, we find that there is an initially steep drop-off in $t_c$, and that $t_c$ continues to fall with $U$ until it reaches the limit set by our image acquisition rate, as shown in \figname{\ref{f_ContactTime}(c)}. The standard deviation of contact formation times, $\sigma_{t_c}$, similarly falls off rapidly as $U$ increases, as shown in \figname{\ref{f_ContactTime}(d)}.  

The linear stability of a viscous film between a liquid and a solid surface subject to destabilizing interfacial forces\citpname{Droplet:Israelachvili2011} and surface tension has been evaluated in prior work\citpname{Droplet:Wyart1990}, and film stability has been probed experimentally for polymer films on silicon wafers\cite{reiter1999thin}. In this analysis, a timescale for linear instability $\tau$ is given in terms of the interfacial force parameter given by the Hamaker constant $A$ and the wave-number $k$ of the instability as
\begin{equation}
\frac{1}{\tau} = -\frac{k^2 h^3}{3 \eta} \left(  \gamma k^2 + \frac{A}{2 \pi h^4} \right).
\label{e_stability} 
\end{equation}
Here, $k = 2 \pi / \lambda$ is the wave-number of the instability, where $\lambda$ is the wavelength; $\eta$ is the air viscosity; $\gamma$ is the liquid surface tension; and $h$ is the steady-state film thickness.
In order to compare our measurements with the theory, we must postulate values for the Hamaker constant and the wave-number. In the absence of a directly measured Hamaker constant for silicon oil and mica across air, we take the measured value of the Hamaker constant for mica interacting with mica across air, calculated with Lifshitz theory to be $-1e-19$ J, and measured to be $-1.35e-19$ J\citpname{Droplet:Israelachvili2011}. The value of $A$ and $K$ that provide best agreement with the observed boundary between spontaneous and nucleated contacts is estimated to be $A = -3e-19$ J and $\frac{k}{2 \pi} = 2e5$ m$^{-1}$, where $A$ is within a factor of about 2 of the measured value of $A$ for mica-air-mica interfaces. Using these values, we compare $t_c$ as a function of $h_{\min}$, substituting $t_c$ for $\tau$ in the prediction of linear instability, as shown in \figname{\ref{f_ContactTime}(e)}.

\section{Discussion}
We present calibrated measurements of the nanometer-scale air film that forms prior to liquid solid contact during droplet impact. With our measurements, we identify two distinct forms of contact initiation - nucleated, and spontaneous - that are distinguished by the relative location of the localized wetting bridge and the location where the air film is most thin. For nucleated contacts, there is no correlation between these locations, whereas for spontaneous rupture events, there is nearly exact coincidence between the location of contact formation and the thinnest region of the air film. Experiments were conducted with silicon oil and water-glycerol mixtures, and the surface of impact is carefully cleaved atomically smooth mica. Water-glycerol solution droplets tend to nucleate contact from greater heights than silicon oil droplets; however, both form contact spontaneously from a narrow distribution of film thicknesses, ranging from a minimum of 18 to a maximum of approximately 30 nanometers, with a measurement error of less than 8 nm. The statistics of contact are consistent with a spinodal dewetting mechanism whereby the air film is destabilized by interfacial attraction of the liquid to the solid surface, as demonstrated by the timescale and height observed for spontaneous and nucleated contact formation.

A natural question arises in considering the stability of the air film - what possible mechanisms could cause the air film to rupture, and liquid-solid contact to form at a discrete point? Mechanical disruption of the air layer is the most common cause of contact initiation on glass surfaces, but for atomically smooth mica, this is no longer the case. When mechanical nucleation sites are not present, other mechanisms become dominant. A few candidate mechanisms include gas kinetics / finite Knudsen number effects, electrostatic charge, and dipole-dipole interactions typically called van der Waals forces. These have been analyzed or explored in prior works, both theoretically\cite{Droplet:Chubynsky2020} and experimentally\cite{Droplet:Gao2019+}; however, controlled experiments with calibrated measurement of the air film thickness have been lacking. With relatively small error bars, and repeated experiments, we identify the tightly-bound range of heights immediately prior to local destabilization of the air film, and furthermore identify parameter values that are consistent with van der Waals attraction as the dominant destabilizing mechanism. Gas kinetic parameters may be responsible for accelerating these dynamics\citpname{Droplet:Chubynsky2020}, as we notably identified a larger estimated value of the Hamaker constant than has been measured for similar material systems\cite{Droplet:Israelachvili2011} to obtain agreement with the observed boundary between spontaneous and nucleated contact formation. Interestingly, the wave length that generates the best agreement with our experimental data is 5 microns, as shown in \figname{\ref{f_ContactTime}(d)}; given the importance of viscosity in coalescence\cite{burton2007role, eggers1999coalescence, paulsen2011viscous}, it is possible that this wave length is somehow set by the viscosity of the liquid; however, to address this question requires further investigation, and is beyond the scope of the current work.

The transition from droplet rebound to spontaneous contact is immediate for silicon oil - at the critical rebound velocity, the oil drop may bounce or create a ring of contact, as shown in \figname{\ref{f_MinHeight}(c)} . The water-glycerol droplets behave differently. There is a range of $U=0.55-0.7$ m/s, wherein no rebound or spontaneous contacts were observed - all the contacts were nucleated. Despite several tries with freshly made solutions, use of different mica sheets, and change of nozzle, we did not observe \emph{any} spontaneous contact formation for this range of impact velocities of solution impact, as can be seen in \figname{\ref{f_Nucleation}(b)}. What could possibly cause the discrepancy between the silicon oil and the water-glycerol solution, given that the mica sheets are identically prepared, and great care was taken to ensure liquids were devoid of dust and dirt? One possibility is that the water-air interface became charged moreso than the silicon oil-air interface. Indeed, if one supposes that randomly fluctuating electromagnetic fields exist in the laboratory, generated by \eg~cameras, the laser, etc. that could not be isolated electrically from the impact surface, then the large dielectric constant of the water\citpname{Droplet:Behrends2006} may cause the water-air interface to charge \emph{more} than the silicon oil-air interface. Indeed, pure silicone oil droplets (without water absorption) do not contain a sufficient concentration of any ions and counter-ions for the appearance of any Coulomb force or Maxwell pressure even in strong electrical fields\citpname{Droplet:Granda2022}. As the mica was prepared in the same way for each experiment, we assume a similar static charge distribution on the mica surface\citpname{TIR:Christenson2016}. Thus, it is possible that the water-air interface contained some static charges that trigger nucleated contact for an intermediate range of impact velocity, where the timescale for spontaneous rupture of the air film was too long to be observed before nucleation took over. This is supported by the large height values of nucleated contacts for the water-glycerol solution droplets, as described in detail in \appname\ref{s_Nucleation}. 

Our results suggest a universal mechanism for spontaneous contact formation through intervening air - namely, spinodal dewetting of the air film once its thickness reduces to within approximately 20 nm. Electrostatics may disrupt this process. Such robustness of the thickness scale for silicon oil suggests that numerical calculations might simply assume contact at this distance, reducing the requisite resolution for accurate numerical modeling of contact phenomena. How this observation is affected by air pressure or applied fields remains to be determined. 

\figuresectionname


\appendix

\clearpage

\section{Microscopy Setup and Calibration}
\label{s_Calibration}
\begin{figure}[!t]
 \centering
\includegraphics[width=0.99\textwidth]{\datafoldername/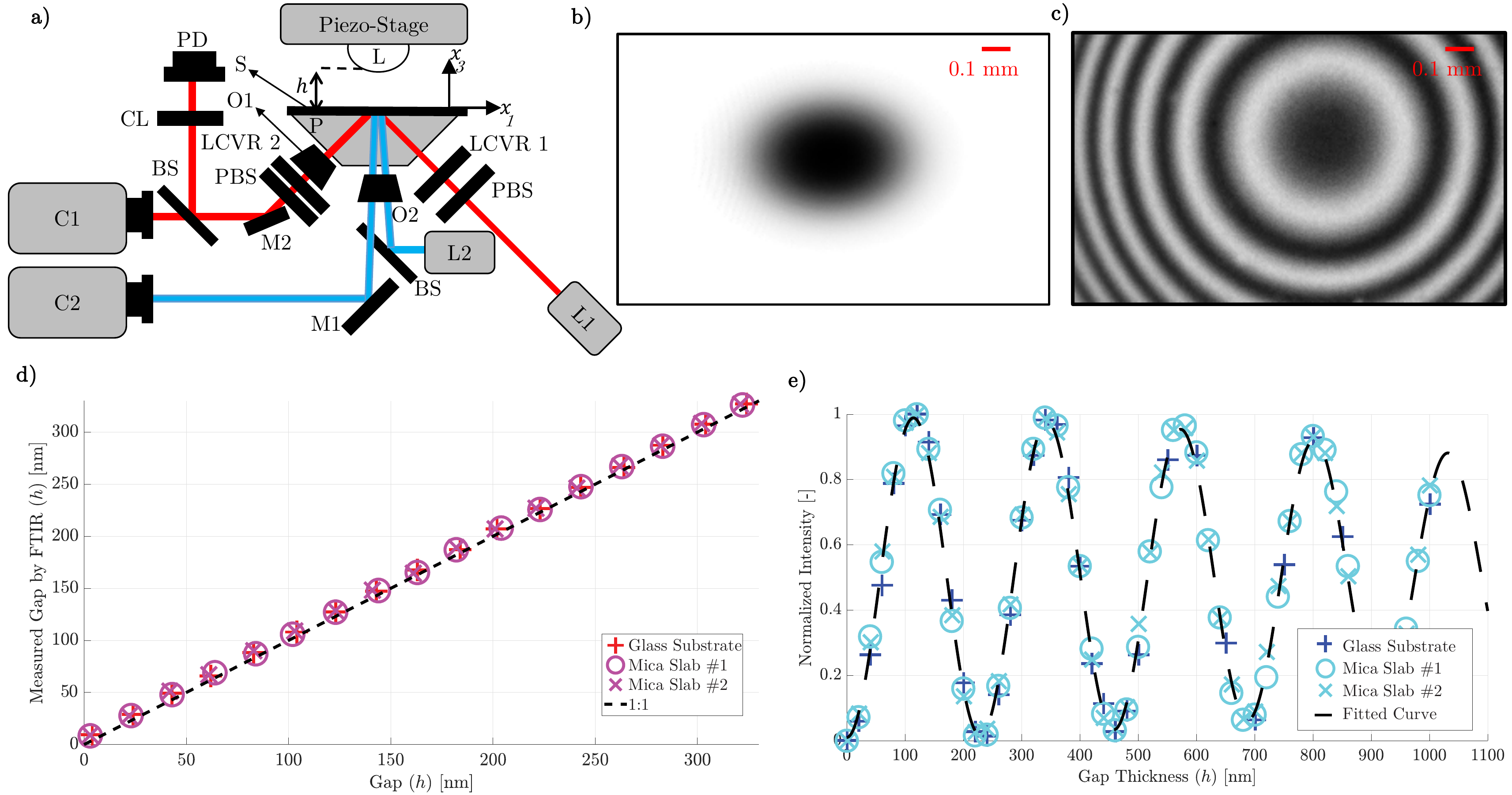}
\caption{
a) The setup used for calibration is similar to that of \figname{\ref{f_DropletSetup}(a)} with the nozzle/luer setup replaced by nm-precise piezeo-stage (Newport 3-axis Piezo stage with strain gage feedback control). The spherical Lens (L) made out of smooth n-BK7 glass is glued to the stage face. We lower the stage with steps of $\Delta x_3=20$ nm, quasi-statically, until the contact occurs and the lens bottom touches the substrate. We calibrate the setup for both mica and glass as solid substrate (S), however, we only used mica for the droplet impact studies reported in this \manuscriptypename. The normalized TIR image and the FIF image at the last step are shown in (b,c), respectively. d) The TIR microscopy is accurate with maximum error of $7$ nm in the range of $0-350$ nm irrespective of the substrate used. e) The universal look-up table for our light source and FIF microscopy setup is applicable to different substrates that we examined as long as the captured images are are normalized with respect to the dark and bright field\citname{FIF:Radler1992}. The horizontal axis of (d,e) are the readings from the piezo stage. 
} 
\label{f_CalibrationCurve}
\end{figure}

The air film thickness measurements are validated by calibration carried out with the simultaneous TIR and FIF microscopy methods. The optical configurations for both methods are depicted in \figname{\ref{f_CalibrationCurve}(a)}. Solid substrates are optically coupled with immersion oil (Nikon) to top of a dove prism (Thor labs, BK-7 glass).

In the TIR method\citpname{Neural:Rubinstein2004,Droplet:Kolinski2012,Droplet:Kolinski2014,Droplet:Kolinski2014+}, the impact surface is illuminated from below with a $5$ mW laser with wavelength $\lambda_{\text{TIR}} = 635$ nm (Arima lasers). The laser is mounted in a temperature controlled mount (model TCLDM9, Thorlabs). The collimated laser light is aligned to be incident at an angle greater than the critical angle for total internal reflection at a substrate-air interface, but smaller than the critical angle at the substrate-liquid interface. The reflected light is imaged by a long working distance objective (Mitutoyo) and tube lens (Thorlabs) onto the sensor of a high speed camera (Photron Fastcam Nova S12). After normalization by the background light intensity, the image intensity is directly related to the air layer thickness by a deterministic transfer function\citname{TIR:Shirota2017} provided the light at the substrate-air interface is linearly polarized. Because we often use mica as the substrate, and mica is weakly birefringent, we must pre- and post-compensate the polarization of the light. This is achieved using two polarizing beam splitters (Thorlabs), and two liquid-crystal variable retarders (LCVRs, Meadowlark), which can programmatically control the polarization state of the light. 

Notably, the optical properties of mica sheets may differ from one experiment to the next; thus, we incorporate a photodiode (model S5792, Hamamatsu) aligned at the output of the second polarizing beam splitter. The intensity measurements made with this photodiode are used to determine the unknown parameters in our optical setup. We introduce $\mathfrak{n}_\text{G}$, $\mathfrak{n}_\text{S}$, $\phi_0$, $\Phi_1$ and $\Phi_2$, and $\mathcal{E}$ as refractive indices of air (gas), solid (substrate), the light angle of incident ($= 45$ deg), the retardance angles of the 2 LCVRs, and the light electric field, respectively. In TIR with linearly polarized light, for a given weakly birefringent substrate with an arbitrary optical axis (${\Psi}_\text{S}$), extended Jones matrix algebra\citpname{TIR:Gu1993} gives the electric field vector ($\mathcal{E}$) as the known functions:
\begin{equation}
\mathcal{E}^{\text{out}} = [\mathcal{E}^{\text{out}}_{\bot}, \mathcal{E}^{\text{out}}_{\|}]^{T} = F \Big(  \mathfrak{n}_\text{G}, \phi_0, \Phi_1, \Phi_2,\mathfrak{n}_\text{S},  {\Psi}_\text{S},  \mathcal{E}^{\text{in}}  \Big),
\label{e_FTIR_Jones2}
\end{equation}
\begin{equation}
\mathcal{E}^{\text{SG}} = [\mathcal{E}^{\text{out}}_{\bot}, \mathcal{E}^{\text{out}}_{\|}]^{T} = G \Big(  \mathfrak{n}_\text{G}, \phi_0, \Phi_1, \mathfrak{n}_\text{S},  {\Psi}_\text{S},  \mathcal{E}^{\text{in}}  \Big),
\label{e_FTIR_Jones1}
\end{equation}
where the superscripts `in', `out' and `SG' specify the input, the output as measured at the photodiode, and the substrate-air interface, respectively; additionally, the subscripts ${\bot}$ and ${\|}$ denote the  s-polarized  and p-polarized  components of the light electric field, respectively.
The two polarizing beam splitters in our setup ensure that the s-polarized portion of the electric field, 
\begin{equation}
\mathcal{E}^{\text{in,out}}_{\bot} = 0.
\end{equation}
The photo diode measures theoretical $|\mathcal{E}^{\text{out}}_{\|}|^2$, quantified by $|\mathcal{E}^{\text{PD}}_{\|}|^2$. The values of $\mathfrak{n}_\text{S}$ and ${\Theta}_\text{S}$ are not known a-priori, and measuring them using a refractometer and polarimeter for each substrate is not feasible for repeated studies with freshly cleaved surfaces required of our study of droplet impacts. 

To overcome the uncertainty of the orientation of the weakly birefringent mica surface in a given iteration of our experiment, we establish the following procedure to determine the LCVR settings that ensure $\mathcal{E}^{\text{SG}}_{\bot} = 0$, using the following in-situ measurement protocol:

\textbf{Step 1:} We introduce the space $x=[\mathfrak{n}_\text{S},\phi_0,{\Psi}_\text{S},\mathcal{E}^{\text{in}}_{\|}]$ that is fully-determined by a system of four equations; however, to increase the accuracy and robustness of this procedure given the uncertainty of the measured data, we solve for $x^\text{opt}$ in an optimization problem: instead of solving four equations, we minimize $(\mathcal{E}^{\text{out}}_{\|}-\mathcal{E}^{\text{PD}}_{\|})^2$ by sweeping a range of voltages on the LCVRs on 25 discrete points, where the retardations of LCVRs 1 and 2 are set independently among $\frac{\Phi_1}{2\pi}=0.25,0.3,0.4,0.5,0.6$ and $\frac{\Phi_2}{2\pi}=0.2,0.3,0.4,0.5,0.55$.

\textbf{Step 2:} In this range of retardation values realized in the LCVRs, the function $F$ has the highest sensitivity to the values of $\mathfrak{n}_\text{S}$ and ${\Theta}_\text{S}$. The root of $F$ is solved for the desired value of $x^\text{opt}$ using a non-linear least-squares optimization routine. To check whether the protocol ran successfully, we verify $\phi_0^\text{opt} = 45$ deg, as is set physically in the optical path. At the end of this step, $\mathfrak{n}_\text{S}$ and ${\Theta}_\text{S}$ are known.

\textbf{Step 3:}  We numerically solve \eqname{\ref{e_FTIR_Jones2}} for $\Phi_1^\text{target}$ that  removes all the unwanted spill ($\frac{\mathcal{E}^{SG}_{\bot}}{\mathcal{E}^{SG}_{\|}} \rightarrow 0$) at the substrate-air interface. From \eqname{\ref{e_FTIR_Jones1}}, the value of $\Phi_2^\text{target}$ that maximizes $\mathcal{E}^{\text{out}}$ is determined; this ensures we maximize the dynamic range captured on the camera's imaging sensor.  

\textbf{Step 4:} The values of $\Phi_1^\text{target}$ and $\Phi_2^\text{target}$ are recorded for each substrate, and the corresponding voltage amplitudes are applied to LCVRs.

Calibration of the microscopy techniques employed in this work is essential. To calibrate the measurements directly, we used the same optical elements of \figname{\ref{f_DropletSetup}(a)} to track a prescribed trajectory of a rigid glass lens surface in the field of view. To prescribe the trajectory with sufficient accuracy, a piezo stage with nm precision and position feedback via strain gauges (NPC3SG, Newport) was used, as shown in \figname{\ref{f_CalibrationCurve}(a)}. A fixed reference geometry in the form of a spherical glass lens (Thorlabs, $f = 0.25$ m) was affixed to the stage surface, and swept step-wise in the negative $x_3$ direction starting from $h \approx350$ nm with a step size of $20$ nm until the lowest points of the lens touched the substrate. At each step of piezo stage, the synchronized cameras record 250 FIF and 250 TIR images at $125$ FPS. Vibrational noise was reduced by rigid mounts and sorbothane sheets inserted between the stage and the optical surface. Despite these measures to control unwanted vibration, mechanical noise with amplitude of $\approx 20$ nm remained; to filter this noise, we took the median of the recorded images at each step of the piezo. The position of the stage at which contact occurs is evident from the lens profile recorded by the cameras (\figname{\ref{f_CalibrationCurve}(b) \& (c)}). Upon contact initiation, mechanical vibrations are no longer measurable. For calibration redundancy, hundreds of control points on the glass lens surface were probed in each image.

The TIR images were re-sized in the $x_2$ direction, and the intensity was mapped to the gap as described elsewhere\citpname{TIR:Shirota2017,Neural:Chantelot2021}. The calibration procedure was repeated for different mica sheets and glass substrates; the maximum absolute error in profilometry was found to be less than $7$ nm for gaps from $h=0-350$ nm, as shown in \figname{\ref{f_CalibrationCurve}(d)}. 

The FIF imaging\citname{FIF:Radler1992} is conducted along a second optical path with a single color LED as shown in \figname{\ref{f_CalibrationCurve}(a)}. The FIF optical path is oriented perpendicular to the substrate surface. Unpolarized light from from a high power blue LED (ILS Solutions) with a peak wavelength of $455$ nm is collimated and directed onto the optical path using a 50-50 beam splitter. The light is then and focused on the substrate surface by a long-working distance microscope objective (Mitutoyo 5X). The interference between the two reflected beams from the varying gap results in a pattern of fringes. The fringes are imaged onto the sensor of a second high-speed camera (Photron S12).

The FIF method can extend the depth-of-field of the gap profilometry to gaps approaching several microns\citname{FIF:Radler1992}. The mapping function from intensity measurements carried out by the monochromatic FIF modality generates uncertainties in the interpretation of the interferogram due to a finite angle of incidence of the light, and the finite light source spectrum\citname{FIF:Radler1992,FIF:Schilling2004,DynamicContact:deRuiter2015+}. To address these sources of uncertainty, we recorded images of a known geometry attached to a nanometer-precise piezo driven stage, shown in \figname{\ref{f_CalibrationCurve}(e)} . The FIF images are normalized with respect to the dark and bright fields\citname{FIF:Radler1992}. We found the mapping function to be universal, and fits any point trajectory on the calibration lens even by changing the substrate with an accuracy of +$8/-3$ nm.

\section{Nucleation of Contact}
\label{s_Nucleation}

Nucleated contacts are distinct from spontaneous contacts due to the random and isolated locations at which they form. This is all the more apparent in the uniformity of the distribution of nucleated contact heights as a function of impact velocity - rather than systematically decreasing, as the globally minimal film thickness $h^{*}$ does for impact velocities below the bouncing transition impact velocity, nucleated contacts can be observed at heights significantly larger than $h^{*}$ for a given impact event, even in cases where spontaneous contacts would otherwise form. 

\begin{figure}[!t]
 \centering
\includegraphics[width=0.9\textwidth]{\datafoldername/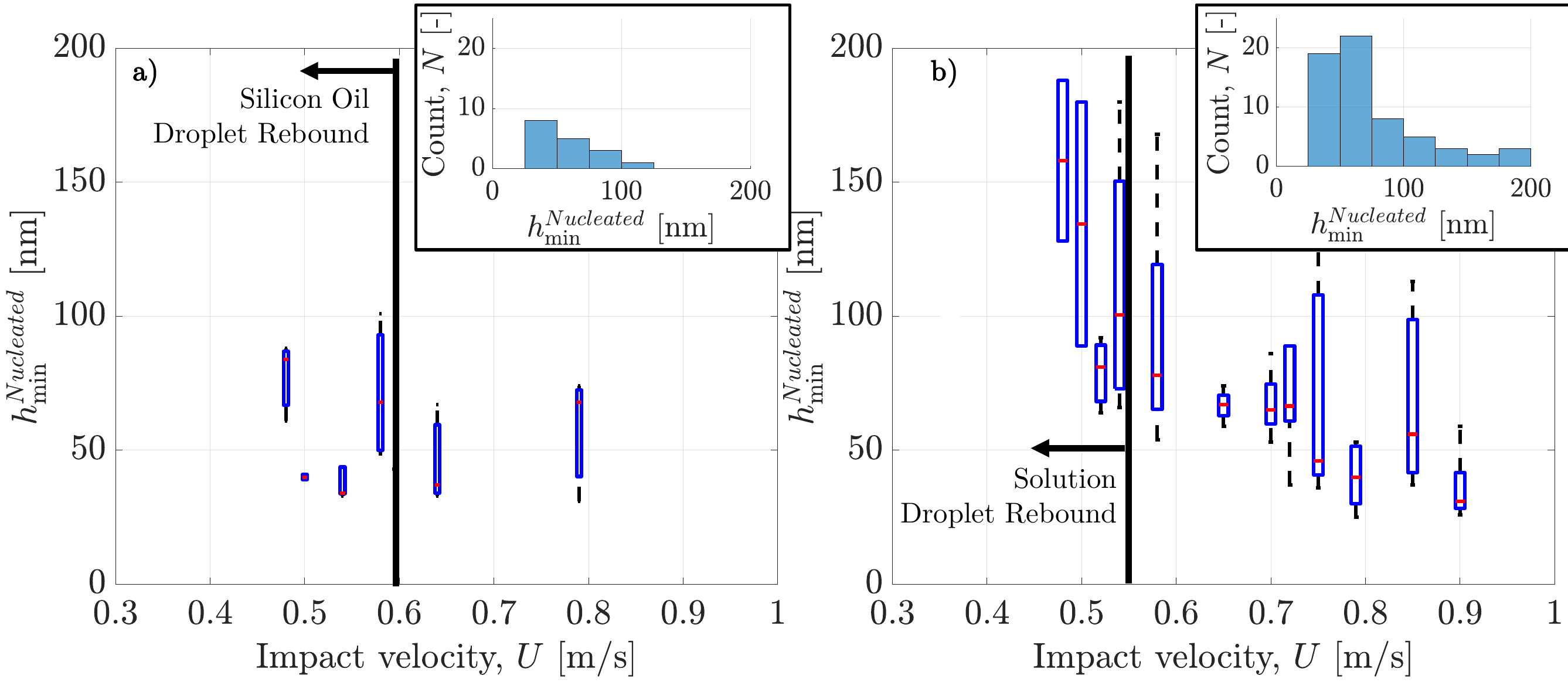}
\caption{
For nucleated contact formation between the liquid and the solid, the lowest film heights $h_{\min}$ are shown for a) the silicon oil and b) the water-glycerol solution. The distribution of $h_{\min}$ appears to be randomly distributed, albeit with much greater nucleation heights for the water-glycerol droplets for impacts below the threshold rebound velocity.  
Out of 91 impact experiments with silicon oil, 38 droplets bounced; 36 droplets made spontaneous and 17 droplets made contact via nucleation. In the experiments with the solution droplets, from total 93: only 15 droplets bounced; 18 droplets made spontaneous contact, and 60 made contact via nucleation.  
} 
\label{f_Nucleation}
\end{figure}

To characterize the nature of nucleated contacts, we plot $h_{\min} (U)$ for the silicon oil and the water-glycerol solution in \figname{\ref{f_Nucleation}(a) \& (b)}, respectively. Whereas the distribution of nucleated contacts for the silicon oil is heavily weighted toward greater proximity to the surface, the same is not true of the contacts formed by impacting droplets of the water-glycerol solution, which are observed at a far greater mean height, with much greater spread, as can be seen in \figname{\ref{f_Nucleation}(a) \& (b), inset}. The difference is unlikely to be exclusively dominated by the different nature of mica surfaces, which for all experiments are prepared identically; rather, some other mechanism of contact formation may be at play for contacts between the water-glycerol solution and the mica. Electrostatic charges might be responsible for the observed difference, because of the different dielectric properties of the silicon oil and water-glycerol solution, but tests of droplet charging in the ambient environment were inconclusive.   

\section{Stability of the Air Film}
\label{s_Stability}

The liquid-air interface beneath the droplet, by nature of the interfacial attraction between the liquid and the solid, is inherently unstable. Indeed, there is a strong attraction between the liquid and the solid, driven by interfacial forces, which are in `competition' with surface tension - that acts to keep the interface flat. 

\begin{figure}[!t]
 \centering
\includegraphics[width=0.85\textwidth]{\datafoldername/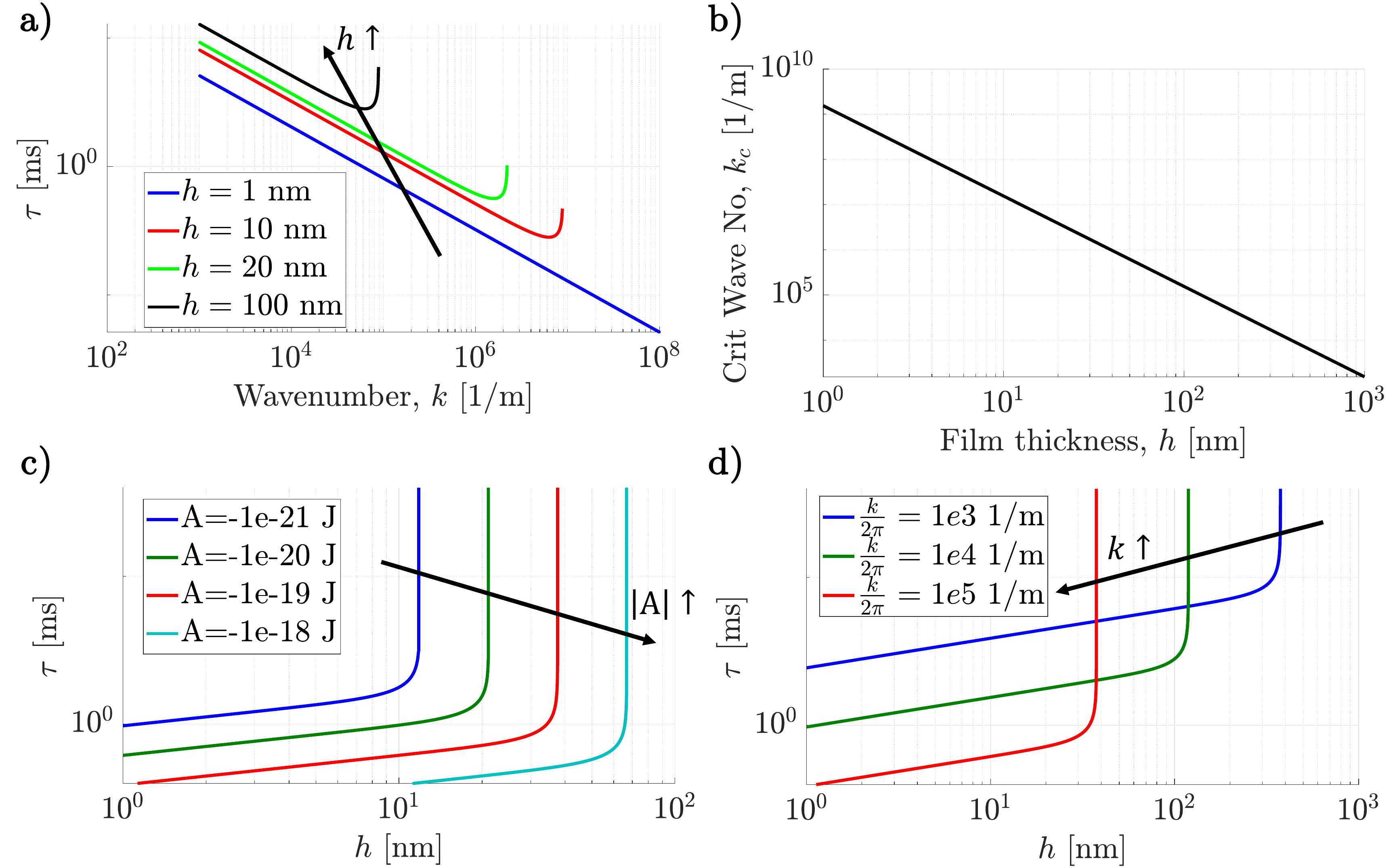}
\caption{
Results of linear stability analysis. a) The critical time, $\tau$, at which the air film loses stability is plotted as a function of wave-number $k$ values of the film thickness $h$, as shown in the colored lines labeled in the legend. b) The critical wave-number $k_c$ scales inversely as the square of $h$ for fixed values of the Hamaker constant $A=1e-19$ J and silicon oil surface tension $\gamma$. The behavior of $\tau$ with respect to $h$ is shown by the family of curves in (c) \& (d). For a fixed $k=2\pi \times 1e5$ 1/m, the curves shift to right and downward with increasing the magnitude of the Hamaker constant; while they move to left and down by increase in $k$ for a constant $A=1e-19$ J.
}
\label{f_Stability}
\end{figure}

From \eqname{\ref{e_stability}}, the stability of the air film can then be assessed for different values of $h$ and $k$. The time constant for linear instability is seen to decrease strongly as a function of wave-number for several different values of $h$, as shown in \figname{\ref{f_Stability}(a)}. For fixed $A$ and $\gamma$, the critical wave-number is a strong function of the film thickness, and scales as $k_c \propto 1/h^2$, as can be seen in \figname{\ref{f_Stability}(b)}.

The stability boundary presented in \figname{\ref{f_ContactTime}(e)}  was generated using estimated values of $A$ and $k$. To establish the effect of $A$ and $k$ on the stability boundary in $t_c$ - $h_{min}$ space, we calculated first-passage time predictions for different values of $A$ and $k$; notably, increasing $A$ magnitude pushes the stability boundary toward the lower right of the graph, while increasing $k$ pushes the stability boundary toward the lower left of the graph, as can be seen in \figname{\ref{f_Stability}(c) \& (d)}, respectively. 

\bibliography{References}

\end{document}